\documentclass{article}

\usepackage{graphicx}

\font\tbff=cmbx10 scaled \magstep4

\font\tenmib=cmmib10 \textfont"E=\tenmib
\font\tenbsy=cmbsy10 \textfont"F=\tenbsy

\def\be{\begin{equation}}
\def\ee{\end{equation}}
\def\bea{\begin{eqnarray}}
\def\eea{\end{eqnarray}}

\title{\large \tbff Eppur si espande }

\author{Marek A. Abramowicz$^{1,2}$, Stanis{\l}aw Bajtlik$^{2}$,\\
Jean-Pierre Lasota$^{3,4}$ \& Audrey Moudens$^{5}$\\
{\small $^{1}$Physics Department, G{\"o}teborg University, SE-412-96 G{\"o}teborg, Sweden}\\
{\small $^{2}$Nicolaus Copernicus Astronomical Center, Bartycka 18, 00-716
Warszawa, Poland}\\
{\small $^{3}$Institut d'Astrophysique de Paris, UMR 7095 CNRS, Universit\'e
P. et M. Curie},\\
{\small 98bis Bd Arago, 75014 Paris, France}\\
{\small $^{4}$Astronomical Observatory, Jagiellonian University,}\\
{\small ul. Orla 171, 30-244 Krak\'ow, Poland}\\
{\small $^{5}$Master 2, Universit\'e de Paris-Sud 11, 91405 Orsay, France}\\
}

\begin{document}

\maketitle

\begin{abstract}
The rather wide-spread belief that cosmological expansion of a flat
3D--space (with spatial curvature $k=0$) cannot be observationally
distinguished from a kinematics of galaxies moving in a flat and
non-expanding space is erroneous. We suggest that the error may have
its source in a non relativistic intuition that imagines the
Universe not as a {\it spacetime} but separates space from time and
pictures the cosmological expansion as space evolving in time. The
physical reality, however, is fundamentally different --- the
expanding Universe is {\it necessarily} a curved spacetime. We show
here that the fact that the {\it spacetime} is curved implies that
the interpretation of the observed cosmological redshift as being
due to the expansion of the cosmological 3D--space is observationally
verifiable. Thus it is impossible to mimic the true cosmological redshift
by a Doppler effect caused by motion of galaxies in a non-expanding 3D-space,
flat or curved. We summarize our points in simple spacetime diagrams that
illustrate a gedanken experiment distinguishing between expansion of
space and pure kinematics. We also provide all relevant mathematics.
None of the previously published discussions of the issue, including
a recent popular Scientific American article
\cite{ScientificAmerican}, offered a similarly clear way out of the
confusion.
\end{abstract}

%%%%%%%%%%%%%%%%%%%%%%%%%%%%%%%%%%%%%%%%%%%%%%%%%
\section{Introduction} \label{section-introduction}
%%%%%%%%%%%%%%%%%%%%%%%%%%%%%%%%%%%%%%%%%%%%%%%%%
%%%%%%%%%%%%%%%%%%%%%%%%%%%%%%%%%%%%%%%%%%%%%%%%%%%%%%%%%%%%%%%%
\begin{figure}
\begin{center}
\includegraphics[width=\textwidth]{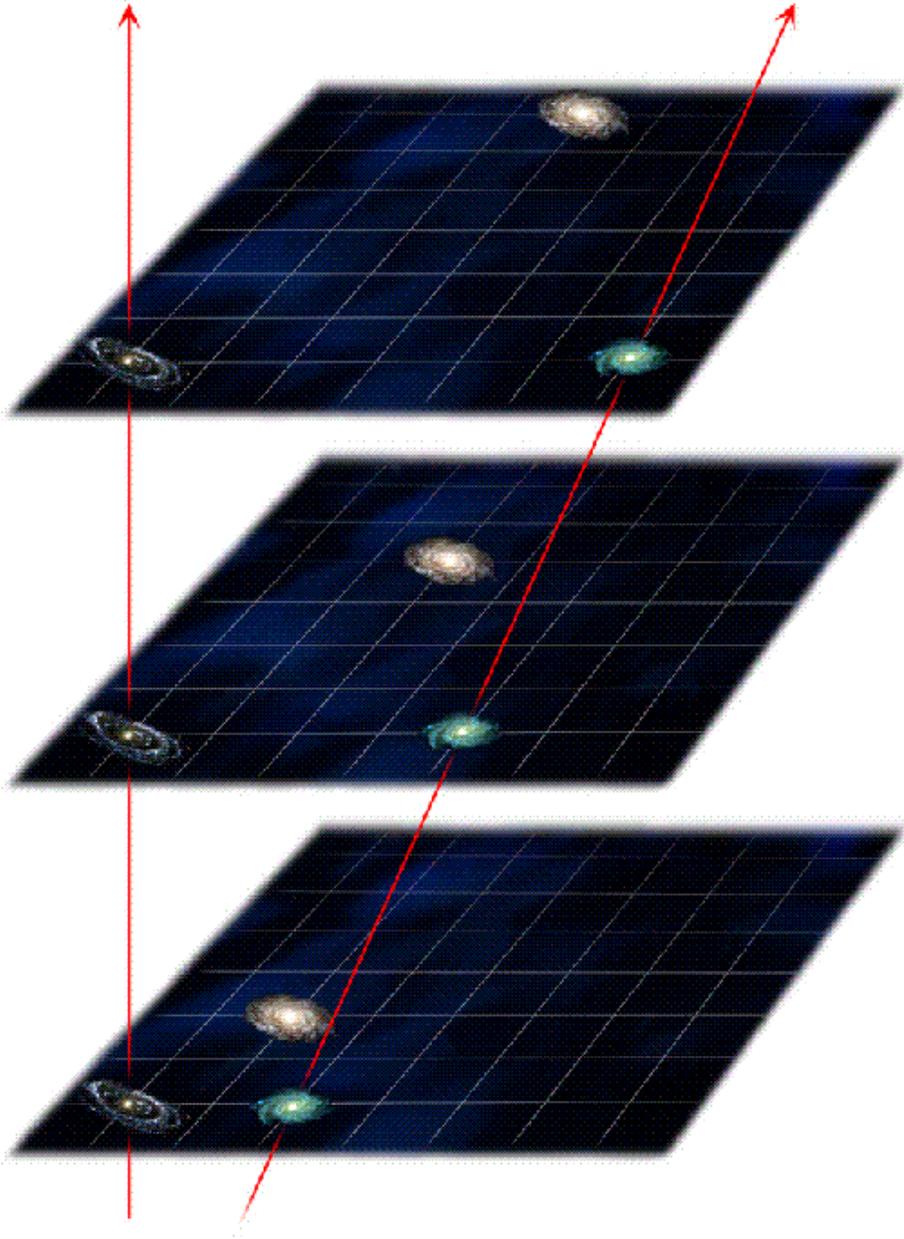}
\caption{Galaxies moving in a non-expanding space}
\label{nexp}
\end{center}
\end{figure}
%%%%%%%%%%%%%%%%%%%%%%%%%%%%%%%%%%%%%%%%%%%%%%%%%%%%%%%%%%%%%%%%%%
%%%%%%%%%%%%%%%%%%%%%%%%%%%%%%%%%%%%%%%%%%%%%%%%%%%%%%%%%%%%%%%%%%
\begin{figure}
\begin{center}
\includegraphics[width=\textwidth]{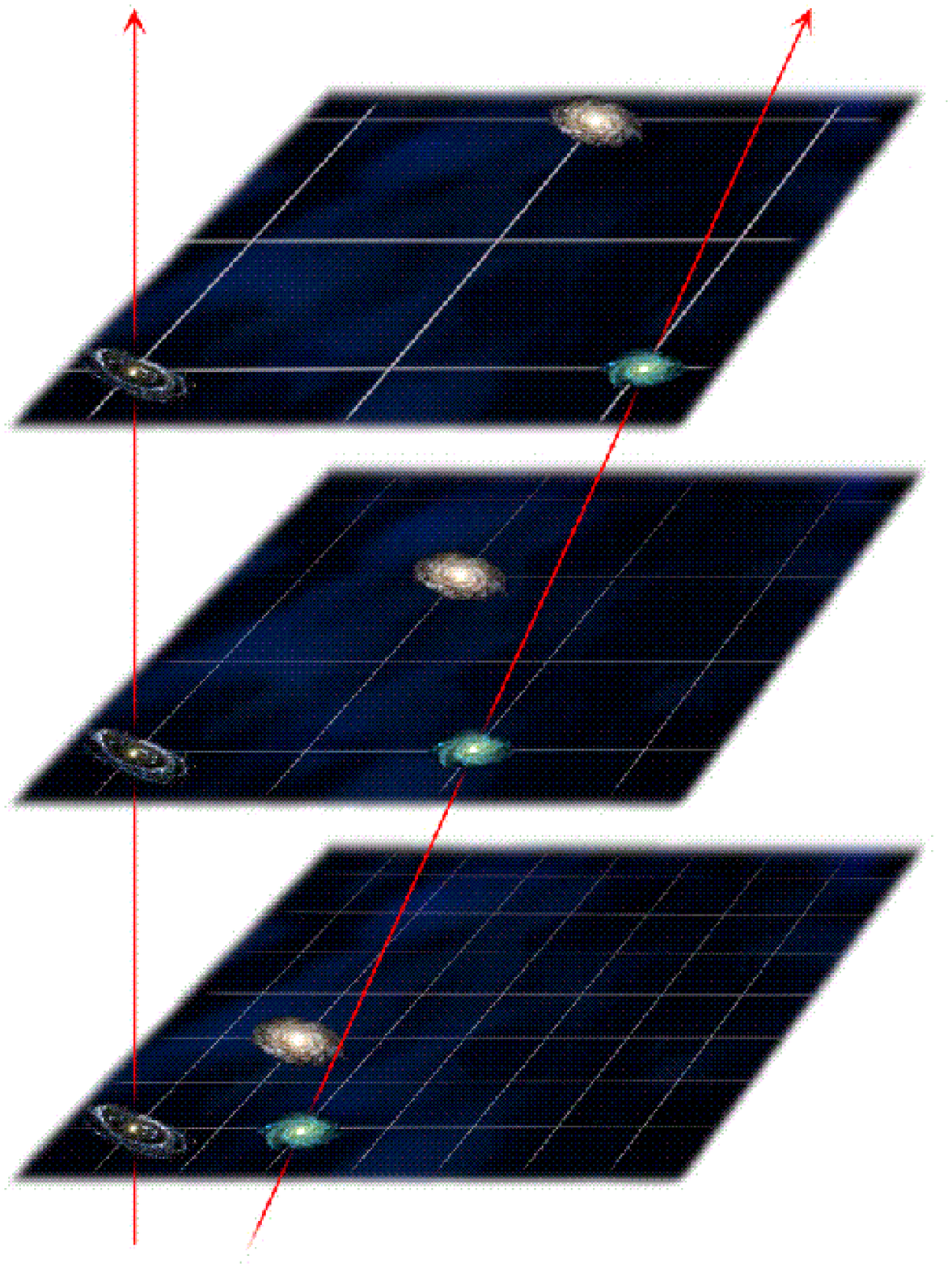}
\caption{Galaxies moving in an expanding space.}
\label{exp}
\end{center}
\end{figure}
%%%%%%%%%%%%%%%%%%%%%%%%%%%%%%%%%%%%%%%%%%%%%%%%%
This work was triggered by skeptical remarks made by Bohdan
Paczy{\'n}ski and helped by several technical discussions with
Micha{\l} Chodorowski. For different reasons (and using different
arguments \cite{Paczynski}, \cite{Chodorowski}) they both advocated
exploring if in the case of a flat cosmological
3D-space ($k=0$) one needs to invoke space expansion since, they argued,
the observed cosmological redshifts can be (formally) interpreted as a
pure Doppler effect due to motion of galaxies in a non-expanding
flat space and no observation can distinguish this from space
expansion. In this paper we show that such interpretation is not
possible. We are particularly interested in the case of an Universe
with flat space sections i.e. with $k = 0$, but we will keep our
discussion general, including also the cases $k = \pm 1$. We are
{\it not} using the $c =1$ convention here, so the light velocity
$c$ appears in all formulae where it should.

\subsection{Expansion or kinematics?}

Figures \ref{nexp} and \ref{exp} describe the source of the
confusion. Let us imagine two galaxies in the expanding Universe.
When the expansion progresses, the two galaxies are just further
apart. The two figures seem to imply that in the case $k=0$ one can adopt
two equivalent interpretations: (a)~the space is expanding,
(b)~galaxies move relative to each other in a non-expanding flat
space. The cosmological redshift in the case $k = 0$ is therefore
explained as the Doppler effect.

\subsection{The metric and curvature of spacetime}

Figures \ref{nexp2} and \ref{exp2} illustrate again the same two
interpretations as shown in Figures \ref{nexp} and \ref{exp}:
galaxies in an expanding universe, and galaxies moving in a
non-expanding space. This time, however, we show why the two
interpretations are {\it in reality}  physically very different.
Indeed, the case of expanding Universe corresponds to a {\it curved}
spacetime. The case of expanding galaxies corresponds to a flat
(Minkowski) spacetime. It is the spacetime curvature that makes the
light trajectories {\it physically different} in the two cases. This
is a reflection of the fact that Newtonian and general-relativistic
cosmology of an isotropic and homogenous Universe are equivalent
only as long as one does not consider the propagation of light.

Figures \ref{nexp2} and \ref{exp2} suggest also that combined
measurements of redshift and distance may together reveal the
physical, observable difference between the two interpretations.
Note, that for these measurements it is sufficient to consider only
radial motions of galaxies and light signals, and ignore angular
$\theta$ and $\phi$ coordinates. This {\it obviously} means that the
curvature of 3D--space is irrelevant to this problem.

The radial part of the metric of the expanding Universe may be
written as,
%%%%%%%%%%%%%%%%%%%%%%%%%%%%%%%%%%%%%%%%%%%%%%%%%
\begin{equation} \label{metric-expanding}
   ds^2 = c^2\,dt^2 - R^2(t)\,dr^2.
\end{equation}
%%%%%%%%%%%%%%%%%%%%%%%%%%%%%%%%%%%%%%%%%%%%%%%%%
For simplicity, we will describe the dimensionless cosmological
scale factor by a power-law function,
%%%%%%%%%%%%%%%%%%%%%%%%%%%%%%%%%%%%%%%%%%%%%%%%%
\begin{equation} \label{scale-factor}
   R(\tau) = \tau^n;~~~ \tau \equiv {t}/{t_*},
\end{equation}
%%%%%%%%%%%%%%%%%%%%%%%%%%%%%%%%%%%%%%%%%%%%%%%%%
where $t_* =$ const is an arbitrary scale; for example it could be $t_* =
100\,{\rm Mpc}/c$.

The only relevant Riemann curvature tensor component, $R_{trtr}$, of the two-dimensional
metric (\ref{metric-expanding}) with the scale factor (\ref{scale-factor}) obeys
\cite{Synge-Schild},
%%%%%%%%%%%%%%%%%%%%%%%%%%%%%%%%%%%%%%%%%%%%%%%%%
\begin{equation} \label{Riemann}
   R_{trtr} \sim \frac{n(n-1)}{(ct)^2},
\end{equation}
%%%%%%%%%%%%%%%%%%%%%%%%%%%%%%%%%%%%%%%%%%%%%%%%%
i.e. vanishes for $n=0$ and $n=1$. In both cases, the spacetime (\ref{metric-expanding})
must be therefore flat i.e. described by the two-dimensional Minkowski metric. Indeed,
$n=0$ corresponds directly to the Minkowski metric,
%%%%%%%%%%%%%%%%%%%%%%%%%%%%%%%%%%%%%%%%%%%%%%%%%
\begin{equation} \label{Minkowski-metric}
   ds^2 = c^2\,dt^2 - dr^2,
\end{equation}
%%%%%%%%%%%%%%%%%%%%%%%%%%%%%%%%%%%%%%%%%%%%%%%%%
while $n=1$ gives the Milne metric,
%%%%%%%%%%%%%%%%%%%%%%%%%%%%%%%%%%%%%%%%%%%%%%%%%
\begin{equation} \label{Milne-metric}
   ds^2 = c^2\,dt^2 - (t/t_*)^2\,dr^2,
\end{equation}
%%%%%%%%%%%%%%%%%%%%%%%%%%%%%%%%%%%%%%%%%%%%%%%%%
which, after the coordinate transformation
%%%%%%%%%%%%%%%%%%%%%%%%%%%%%%%%%%%%%%%%%%%%%%%%%
\begin{equation} \label{Milne-Minkowski-transformation}
   T = t\,\cosh (r/ct_*), ~~~ X = ct\,\sinh (r/ct_*),
\end{equation}
%%%%%%%%%%%%%%%%%%%%%%%%%%%%%%%%%%%%%%%%%%%%%%%%%
takes the form $ds^2 = c^2\,dT^2 - dX^2$, identical with the
Minkowski metric (\ref{Minkowski-metric}). Milne $n=1$ metric
(\ref{Milne-metric}) and Minkowski $n=0$ metric
(\ref{Minkowski-metric}) are thus identical. However, from the $t=
\rm const.$ 3D--space point of view one can conclude that space is
expanding. Of course since in this case the 4D space-time is flat
this is just an illusion. In the following we will show that this
illusion is the source of confusion about the expansion of
cosmological 3D--space. In a curved spacetime expansion is not
illusory as it corresponds to physical observables.

%%%%%%%%%%%%%%%%%%%%%%%%%%%%%%%%%%%%%%%%%%%%%%%%%
   \section{A gedanken experiment} \label{section-experiment}
%%%%%%%%%%%%%%%%%%%%%%%%%%%%%%%%%%%%%%%%%%%%%%%%%

\begin{figure}
  \begin{center}
  \includegraphics[width=\textwidth]{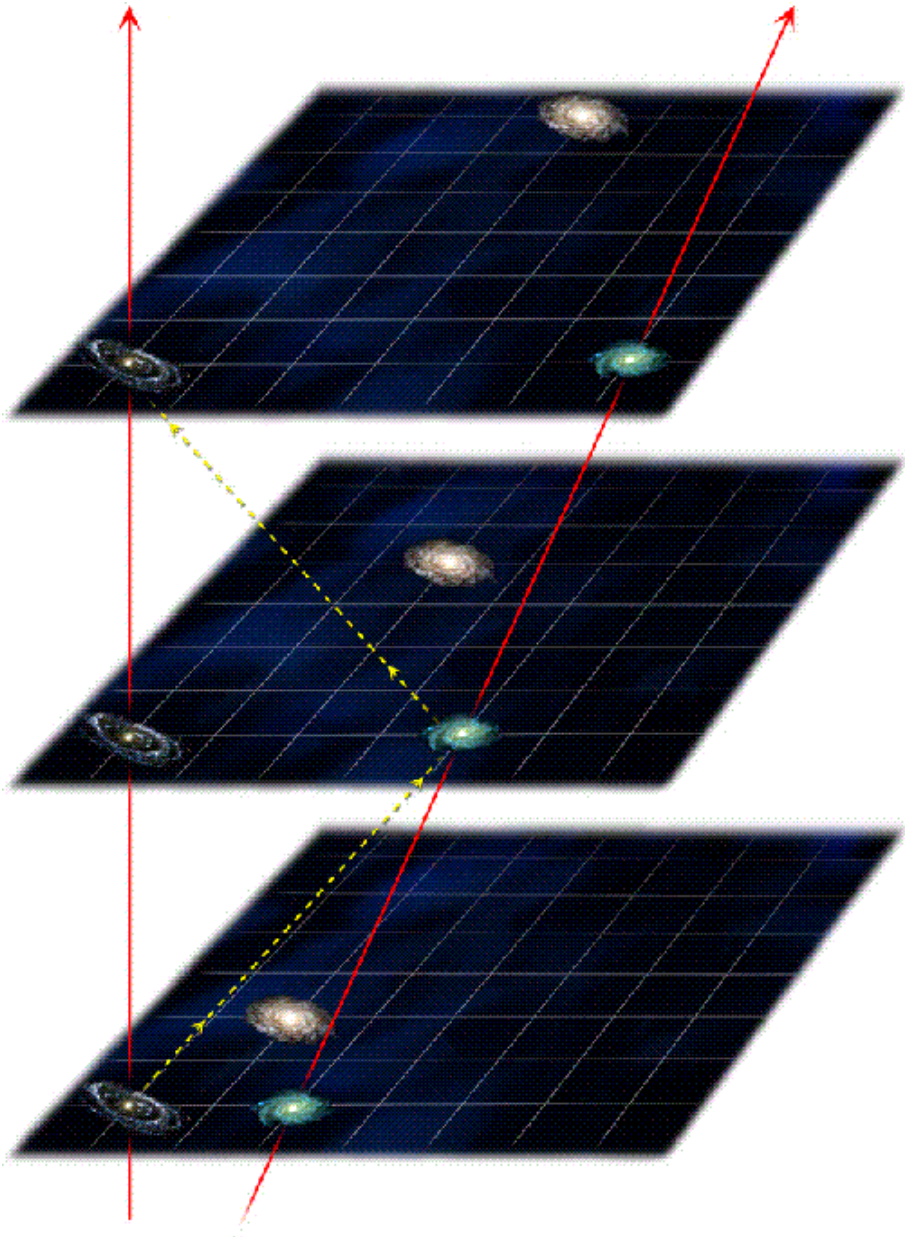}
  \hfill
  \caption{Same as Figure \ref{nexp} but
  with added radar measurement of distance. The light trajectories reveal
  {\it different} spacetime structure from that in Fig. \ref{exp2}(next page).
  Obviously this difference will show up in simultaneous measurements of the
  redshift and distance. Thus, measuring the redshift and
  distance, one can learn whether or not space is really expanding.}
  \label{nexp2}
  \end{center}
\end{figure}

\begin{figure}
  \begin{center}
  \includegraphics[width=\textwidth]{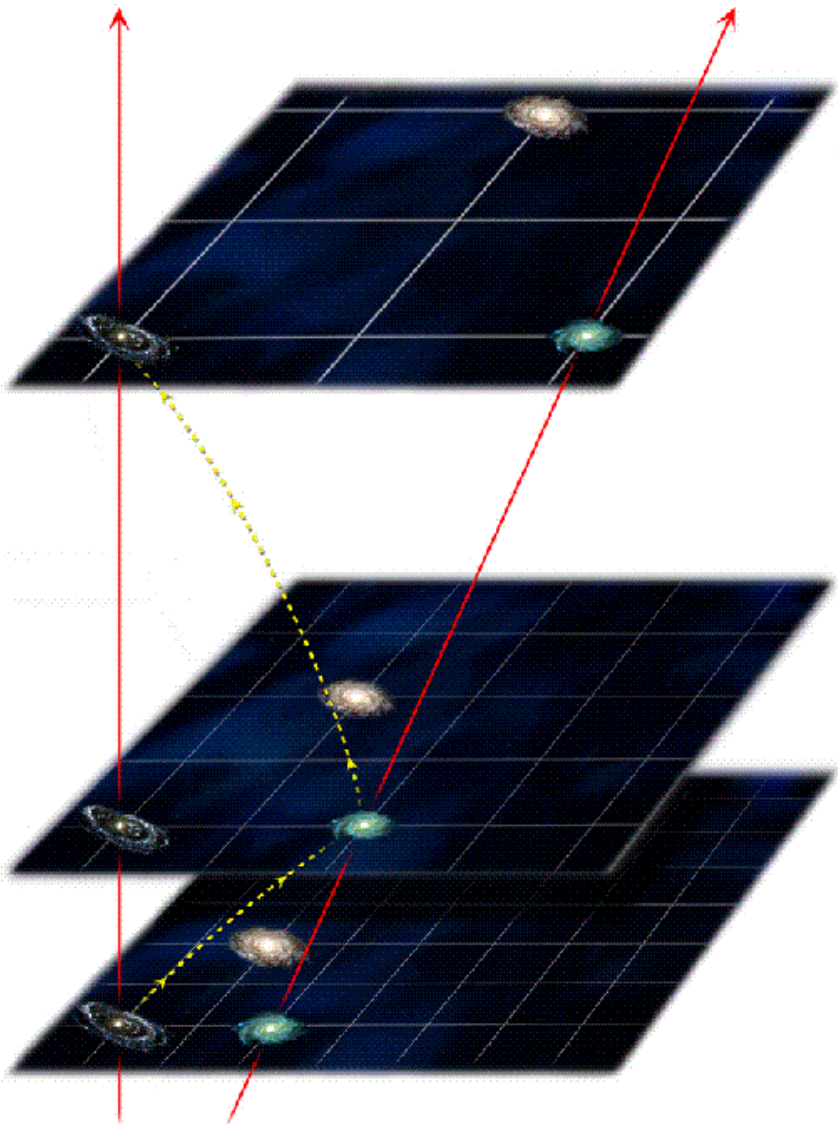}
  \caption{Same as Figure \ref{exp} but
  with added radar measurement of distance. The light trajectories reveal
  {\it different} spacetime structure from that in Fig. \ref{nexp2} (previous page).
  }
    \label{exp2}
  \end{center}
\end{figure}

Figures \ref{nexp2} and \ref{exp2} show the history of a radar measurements of the
distance between two galaxies. At the spacetime event
$[\,t_1,\,0\,]$ a light signal is sent by an observer in the galaxy
located at $r$$=$$0$. At the spacetime event $[\,t_0,\,r_0\,]$ this
signal is reflected from a distant galaxy located in $r$$=$$r_0$ ,
and returns to the observer at the spacetime event $[\,t,\,0\,]$.
The $[\,t,\,r\,]$ coordinates in the metric (\ref{metric-expanding})
are comoving with the galaxies, so that $r$$=$$0$ for the first
galaxy and $r$$=$$r_0$ for the second hold at all times.

The observer continuously measures the pair of moments $t_1$ (at which the signal
is emitted) and $t$ (at which the signal returns) constructing a
function $t_1(t)$. He also continuously records the redshift $z(t)$ of the other galaxy,
and defines and continuously records the ``radar distance'',
%%%%%%%%%%%%%%%%%%%%%%%%%%%%%%%%%%%%%%%%%%%%%%%%%
\begin{equation}
\label{distance-definition}
D_{\rm RAD}(t) = c\, \frac {t - t_1}{2}.
\end{equation}
%%%%%%%%%%%%%%%%%%%%%%%%%%%%%%%%%%%%%%%%%%%%%%%%%

In the remaining part of this article we will discuss these
measurements and show, that from the knowledge of the three
functions, $t_1(t)$, $D_{\rm RAD}(t)$, and $z(t)$, the observer
may definitely decide whether or not the space is expanding. We
stress that the conclusion is based on the results of his gedanken
experiment only, and is independent of the definition of 3D--space
that he, or his colleagues, might adopt.
%%%%%%%%%%%%%%%%%%%%%%%%%%%%%%%%%%%%%%%%%%%%%%%%%
%%%%%%%%%%%%%%%%%%%%%%%%%%%%%%%%%%%%%%%%%%%%%%%%%
%%%%%%%%%%%%%%%%%%%%%%%%%%%%%%%%%%%%%%%%%%%%%%%%%%%%
%%%%%%%%%%%%%%%%%%%%%%%%%%%%%%%%%%%%%%%%%%%%%%%%%%%%%%%%%%%%%%%%%%%%%%%%%%%%%

\subsection{The radar distance in the expanding Universe}

Along the trajectory of a light signal $ds =0$, and therefore one
has,
%%%%%%%%%%%%%%%%%%%%%%%%%%%%%%%%%%%%%%%%%%%%%%%%%
\begin{equation} \label{light-trajectory}
   c\int^{t_0}_{t_1} \frac {dt}{R(t)} = r_0 = c\int^{t}_{t_0} \frac {dt}{R(t)},
\end{equation}
%%%%%%%%%%%%%%%%%%%%%%%%%%%%%%%%%%%%%%%%%%%%%%%%%
or, using (\ref{scale-factor}),
%%%%%%%%%%%%%%%%%%%%%%%%%%%%%%%%%%%%%%%%%%%%%%%%%
\begin{equation} \label{time-sent}
   \tau_1 = \tau\,\left[ 1 - 2\epsilon (1 - n)\,\tau^{-(1-n)} \right]^{1/(1-n)},
\end{equation}
%%%%%%%%%%%%%%%%%%%%%%%%%%%%%%%%%%%%%%%%%%%%%%%%%
%%%%%%%%%%%%%%%%%%%%%%%%%%%%%%%%%%%%%%%%%%%%%%%%%
\begin{equation} \label{time-reflected}
   \tau_0 = \tau\,\left[ 1 - \epsilon (1 - n)\,\tau^{-(1-n)} \right]^{1/(1-n)},
\end{equation}
%%%%%%%%%%%%%%%%%%%%%%%%%%%%%%%%%%%%%%%%%%%%%%%%%
where we introduced the notation
%%%%%%%%%%%%%%%%%%%%%%%%%%%%%%%%%%%%%%%%%%%%%%%%%
\begin{equation} \label{epsilon}
   \epsilon \equiv \frac {r_0}{c\,t_*} = {\rm const}.
\end{equation}
%%%%%%%%%%%%%%%%%%%%%%%%%%%%%%%%%%%%%%%%%%%%%%%%%
We do not assume that $\epsilon \ll 1$.

From (\ref{distance-definition}) and (\ref{time-sent}) it follows that
%%%%%%%%%%%%%%%%%%%%%%%%%%%%%%%%%%%%%%%%%%%%%%%%%
\begin{equation} \label{radar-distance-final}
   D_{\rm RAD}(\tau) = \frac {ct_* \tau}{2} \left[ 1 - \left( 1 -
   \frac {2\epsilon(1-n)}{\tau^{1-n}}\right)^{1/(1-n)} \right].
\end{equation}
%%%%%%%%%%%%%%%%%%%%%%%%%%%%%%%%%%%%%%%%%%%%%%%%%

\subsection{The redshift measurement in expanding universe}

The cosmological redshift measured in time $t$ equals (cf. Figures \ref{nexp2} and \ref{exp2} and
equation (\ref{time-reflected})),
%%%%%%%%%%%%%%%%%%%%%%%%%%%%%%%%%%%%%%%%%%%%%%%%%
\begin{equation} \label{cosmological-redshift}
   1 + z(\tau) = \frac {R(\tau)}{R(\tau_0)} =  \left[ 1 - \frac
   {\epsilon (1 - n)}{\tau^{1-n}} \right]^{-n/(1-n)} \equiv {\cal X}.
\end{equation}
%%%%%%%%%%%%%%%%%%%%%%%%%%%%%%%%%%%%%%%%%%%%%%%%%
Let us define $\delta \equiv 1-n$. Milne's universe correspond
to $\delta = 0$. In this limit, one has 
%%%%%%%%%%%%%%%%%%%%%%%%%%%%%%%%%%%%%%%%%%%%%%%%%
\begin{equation} \label{limit-function-X}
 {\cal X}(\delta \rightarrow 0) = e^{\epsilon}
\end{equation}
%%%%%%%%%%%%%%%%%%%%%%%%%%%%%%%%%%%%%%%%%%%%%%%%%

%%%%%%%%%%%%%%%%%%%%%%%%%%%%%%%%%%%%%%%%%%%%%%%%%
   \section{Is space expanding?} \label{section-redshift}
%%%%%%%%%%%%%%%%%%%%%%%%%%%%%%%%%%%%%%%%%%%%%%%%%

In this Section we calculate the increasing ``redshift distance'' to
a distant galaxy by supposing that the measured redshift may be
interpreted as a pure Doppler effect in a non-expanding space. If
space is not expanding, the redshift distance should be equal to
the measured radar distance (\ref{radar-distance-final}).
If the two distances are not equal one concludes that the space
must be expanding and that this expansion is not equivalent to
recession of galaxies in a static space.

The Doppler interpretation of the redshift assumes that,
%%%%%%%%%%%%%%%%%%%%%%%%%%%%%%%%%%%%%%%%%%%%%%%%%
\begin{equation} \label{Doppler-redshift}
   1 + z(\tau) = \left( \frac {1 + v/c}{1 - v/c}\right)^{1/2}.
\end{equation}
%%%%%%%%%%%%%%%%%%%%%%%%%%%%%%%%%%%%%%%%%%%%%%%%%
Equating $1 + z(\tau)$ in (\ref{cosmological-redshift}) and (\ref{Doppler-redshift}) one calculates the velocity of the 
galaxy $v(\tau)$,
%%%%%%%%%%%%%%%%%%%%%%%%%%%%%%%%%%%%%%%%%%%%%%%%%
\begin{equation} \label{velocity}
   v(\tau)  = c \,\frac {{\cal X}^2 - 1}{{\cal X}^2 + 1}.
\end{equation}
%%%%%%%%%%%%%%%%%%%%%%%%%%%%%%%%%%%%%%%%%%%%%%%%%
From (\ref{limit-function-X}) and (\ref{velocity}) it follows
that in Milne's universe, i.e. for $\delta \rightarrow 0$, 
one has
%%%%%%%%%%%%%%%%%%%%%%%%%%%%%%%%%%%%%%%%%%%%%%%%%
\begin{equation} \label{velocity-Milne-limit}
v(\delta \rightarrow 0) = 
c\,\frac {e^{2\epsilon} - 1}{e^{2\epsilon} + 1} =
c\,\tanh (\epsilon)
\end{equation}
%%%%%%%%%%%%%%%%%%%%%%%%%%%%%%%%%%%%%%%%%%%%%%%%%

The distance to the galaxy, according to the supposed 
interpretation of the redshift as a pure Doppler effect is,
%%%%%%%%%%%%%%%%%%%%%%%%%%%%%%%%%%%%%%%%%%%%%%%%%
\begin{equation} \label{redshift-distance}
   D_{\rm RED}(\tau) = \int_0^{\tau^{\prime}_0} v(\tau) d\, \tau,
   ~~~\tau^{\prime}_0 \equiv \frac {\tau + \tau_1}{2}.
\end{equation}
%%%%%%%%%%%%%%%%%%%%%%%%%%%%%%%%%%%%%%%%%%%%%%%%%

 For $\delta \ll 1$ one has from
(\ref{radar-distance-final}) and from (\ref{redshift-distance}),
%%%%%%%%%%%%%%%%%%%%%%%%%%%%%%%%%%%%%%%%%%%%%%%%%
\bea \label{distances-small-delta}
D_{\rm RAD}(\tau; \delta \ll 1) &=& \frac {ct_*\tau}{2}\left[1 -
   e^{-2\epsilon}\right] \nonumber \\
   &-& \frac {ct_*\tau}{2}\left[ 2\,e^{-2\epsilon}\,(\ln \tau - \epsilon)
   \right]\,\epsilon \delta, \nonumber \\
   D_{\rm RED}(\tau; \delta \ll 1) &=& \frac {ct_*\tau}{2}\left[1
   - e^{-2\epsilon}\right] \nonumber \\
   &+& ct_*\tau \left[ 
   \ln \frac{\tau}{2} + \epsilon ( 2\,\ln \tau - \ln \frac {\tau}{2}
   - 2)
   \right]\,\epsilon \delta 
   .
\eea
%%%%%%%%%%%%%%%%%%%%%%%%%%%%%%%%%%%%%%%%%%%%%%%%%
From the above formula one sees that
%%%%%%%%%%%%%%%%%%%%%%%%%%%%%%%%%%%%%%%%%%%%%%%%%
\bea \label{agreement-disagreement}
   D_{\rm RAD}(\tau) &=&  D_{\rm RED}(\tau)~~{\rm for}~~\delta=0
   \nonumber \\
   D_{\rm RAD}(\tau) &\not =&  D_{\rm RED}(\tau)~~{\rm for}~~\delta\not=0~~(\delta \ll 1).
\eea
%%%%%%%%%%%%%%%%%%%%%%%%%%%%%%%%%%%%%%%%%%%%%%%%%

%%%%%%%%%%%%%%%%%%%%%%%%%%%%%%%%%%%%%%%%%%%%%%%%%
   \section{The Milne Universe} \label{section-milne}
%%%%%%%%%%%%%%%%%%%%%%%%%%%%%%%%%%%%%%%%%%%%%%%%%

In this Section we derive the form of the metric of Milne's Universe in a way that may
be novel to some readers.

The Hubble law says that in the {\it local} Universe the small redshift $z \ll 1$ of a
nearby galaxy is proportional to the galaxy distance, $cz = H{\tilde r}$. For $z \ll 1$,
one may use the non-relativistic formula $z = v/c$ and rewrite the Hubble law as $v =
H{\tilde r}$. Let a galaxy located at a distance ${\tilde r}$ has its velocity $v$. What
would be the velocity of a galaxy at the distance ${\tilde r} + d{\tilde r}\,$? The
spacetime of the Universe is described locally by the Minkowski metric, where
the answer is given by the special relativistic formula for adding velocities,
%%%%%%%%%%%%%%%%%%%%%%%%%%%%%%%%%%%%%%%%%%%%%%%%%
\begin{equation} \label{Lorentz}
   v + dv = \frac {v + Hd{\tilde r}}{1 + vHd{\tilde r}/c^2} = v + Hd{\tilde r}
   \left( 1 - v^2/c^2 \right).
\end{equation}
%%%%%%%%%%%%%%%%%%%%%%%%%%%%%%%%%%%%%%%%%%%%%%%%%
Integrating (\ref{Lorentz}) we get,
%%%%%%%%%%%%%%%%%%%%%%%%%%%%%%%%%%%%%%%%%%%%%%%%%
\begin{equation} \label{Lorentz-integral}
   H\int_0^{\tilde r}\,d{\tilde r} = c\,\int_0^v \frac {d(v/c)}{1-v^2/c^2},
\end{equation}
%%%%%%%%%%%%%%%%%%%%%%%%%%%%%%%%%%%%%%%%%%%%%%%%%
and finally,
%%%%%%%%%%%%%%%%%%%%%%%%%%%%%%%%%%%%%%%%%%%%%%%%%
\begin{equation} \label{Lorentz-final}
   v = c\,\tanh (\frac {H{\tilde r}}{c})
\end{equation}
%%%%%%%%%%%%%%%%%%%%%%%%%%%%%%%%%%%%%%%%%%%%%%%%%
The Hubble constant $H$ may be only a function of time, $H = H(t)$. If the velocity is
constant in time, the distance ${\tilde r}$ must obey ${\tilde r} \sim 1/H(t)$. On the other
hand, one knows that ${\tilde r}(t) = R(t)\,r$, with $r =\,$ const being the comoving
coordinate, and $H(t) = {\dot R}(t)/R(t)$. From these conditions, one concludes that
${\dot R}/R \sim 1/R$, or $R(t) \sim t$. Therefore, in this case, the metric takes the
form
%%%%%%%%%%%%%%%%%%%%%%%%%%%%%%%%%%%%%%%%%%%%%%%%%
\begin{equation} \label{Milne-Milne}
   ds^2 =c^2\,dt^2 - \left( \frac {t}{t_*}\right)^2\,dr^2.
\end{equation}
%%%%%%%%%%%%%%%%%%%%%%%%%%%%%%%%%%%%%%%%%%%%%%%%%
We have arrived again at the Milne metric. Note that the velocity
expressed in terms of the comoving coordinate takes the form that
explicitly shows $v =\,$const,
%%%%%%%%%%%%%%%%%%%%%%%%%%%%%%%%%%%%%%%%%%%%%%%%%
\begin{equation} \label{Lorentz-final-epsilon}
   v = c\,\tanh (\frac{r}{ct_*}) = c\,\tanh (\epsilon),
\end{equation}
%%%%%%%%%%%%%%%%%%%%%%%%%%%%%%%%%%%%%%%%%%%%%%%%%
which agrees with (\ref{velocity-Milne-limit}).

The Milne spacetime is therefore the result of a particular
globalization of the local Minkowskian kinematics.

One can also deduce (\ref{Lorentz-final-epsilon}) from the (already
mentioned) transformation between the Minkowski metric expressed in
terms of $T, X$ and the equivalent Milne's metric expressed in terms
of $t, \epsilon$,
%%%%%%%%%%%%%%%%%%%%%%%%%%%%%%%%%%%%%%%%%%%%%%%%%
\begin{equation} \label{Milne-Minkowski-transformation-epsilon}
   T = t\,\cosh (\epsilon), ~~~ X = ct\,\sinh (\epsilon).
\end{equation}
%%%%%%%%%%%%%%%%%%%%%%%%%%%%%%%%%%%%%%%%%%%%%%%%%
Indeed, from the above equation, it follows directly that,
%%%%%%%%%%%%%%%%%%%%%%%%%%%%%%%%%%%%%%%%%%%%%%%%%
\begin{equation} \label{Lorentz-final-final}
   v = \frac {X}{T} = c\,\tanh (\epsilon).
\end{equation}
%%%%%%%%%%%%%%%%%%%%%%%%%%%%%%%%%%%%%%%%%%%%%%%%%
The full, four dimensional version of the Milne metric that follows from the full four
dimensional Minkowski $ds^2 = c\,dt^2 - dX^2 - X^2\,[d\theta^2 + \sin^2\theta\,d\phi^2]$
after the coordinate transformation (\ref{Milne-Minkowski-transformation},
\ref{Milne-Minkowski-transformation-epsilon}) and $\theta = \theta$, $\phi = \phi$,
has the form
%%%%%%%%%%%%%%%%%%%%%%%%%%%%%%%%%%%%%%%%%%%%%%%%%
\begin{equation} \label{full-Milne-metric}
   ds^2 = c^2\,dt^2 - (t/t_*)^2\,\left[dr^2 + \sinh^2(r/ct_*)\left(d\theta^2 +
   \sin^2\theta \,d\phi^2\right) \right].
\end{equation}
%%%%%%%%%%%%%%%%%%%%%%%%%%%%%%%%%%%%%%%%%%%%%%%%%
The Milne space $t$$\,=\,$const, $dt$$\,=\,$$0$ has the metric
%%%%%%%%%%%%%%%%%%%%%%%%%%%%%%%%%%%%%%%%%%%%%%%%%
\begin{equation} \label{space-Milne-metric}
   d\ell^2 = dr^2 + \sinh^2(r/ct_*)\left(d\theta^2 + \sin^2\theta\,d\phi^2 \right),
\end{equation}
%%%%%%%%%%%%%%%%%%%%%%%%%%%%%%%%%%%%%%%%%%%%%%%%%
and therefore corresponds to the case $k$$\,=$$\,-1$.

The Milne and Minkowski metrics describe {\it the same} non-curved
spacetime of special relativity, however they {\it differently}
divide the spacetime into space and time. In the Minkowski's
interpretation in the $[\,X,\,T\,]$ coordinates, galaxies are moving
with constant velocities $X/T$$\,=\,$$\tanh (\epsilon)$ in space
that is not expanding. In the Milne's interpretation, in the comoving
coordinates $[\,\tau,\,\epsilon\,]$ the space is expanding at the
rate $R(\tau) $$\,=\,$$\tau$, and galaxies have fixed positions
$\epsilon =\,$const in an expanding space. Our condition that
equality of radar and redshift distance measurements indicates a
non-expanding space picks up Minkowski's interpretation as true ---
Milne's interpretation is an illusion, the space of a (4--D) flat Universe
is not expanding.

%%%%%%%%%%%%%%%%%%%%%%%%%%%%%%%%%%%%%%%%%%%%%%%%%
\section{Conclusions} \label{section-conclusions}
%%%%%%%%%%%%%%%%%%%%%%%%%%%%%%%%%%%%%%%%%%%%%%%%%

The concept of a {\it global 3--space} is not fundamental in Einstein's
relativity, and no universal, invariant, definition of global 3--space
is possible. An observer may invariantly define only his {\it local}
3--D space.

In this article we have not attempted to define the global space.
Instead, we showed that independent of the definition, in a
non-expanding space the radar and redshift distances should be the
same. When they are not, one must conclude that space is expanding.
We have shown that if the cosmological spacetime is curved the two
measurements cannot agree. Thus, the expansion of space relates to
the curvature of {\it spacetime}. The curvature of the 3D--space
has nothing to do with it.

M.A.A. and St.B. acknowledge the support from Polish State Committee for Scientific
Research, grants 1P03D 012 26 and N203 009 31/1466, and A.M. acknowledges
the support from the French-Polish LEA-Astro-PF.

\end{document}